\documentclass[acmtog, authorversion]{acmart}

\AtBeginDocument{%
  \providecommand\BibTeX{{%
    \normalfont B\kern-0.5em{\scshape i\kern-0.25em b}\kern-0.8em\TeX}}}

\begin{document}

\title{ProtoFlakes: A Conceptual Modular and Plug-and-Play Prototyping Tool Kit for Smart Jewelry Design Exploration}
\shorttitle{ProtoFlakes}


\author{İhsan Ozan Yıldırım}
\authornotemark[1]
\orcid{0000-0002-2432-147X}
\affiliation{%
  \institution{Arçelik A.Ş. R\&D Sensor Technologies Directorate}
  \streetaddress{Tuzla}
  \city{Istanbul}
  \country{Türkiye}
  \postcode{34950}
}
\email{ihsanozan.yildirim@arcelik.com}

\author{Murat Kuşcu}
\orcid{0000-0002-8463-6027}
\affiliation{%
  \institution{Koç University - Department of Electrical and Electronics Engineering}
  \streetaddress{Sarıyer}
  \city{Istanbul}
  \country{Türkiye}
  \postcode{34450}
}
\email{mkuscu@ku.edu.tr}

\author{Oğuzhan Özcan}
\orcid{0000-0002-4410-3955}
\affiliation{%
  \institution{Koç University - Arçelik Research Center for Creative Industries}
  \streetaddress{Sarıyer}
  \city{Istanbul}
  \country{Türkiye}
  \postcode{34450}
}
\email{oozcan@ku.edu.tr}


\renewcommand{\shortauthors}{Yıldırım, et al.}

\begin{abstract}


The design of smart jewelry can be challenging as it requires technical knowledge and practice to explore form and function. Adressing this issue, we propose ProtoFlakes, a design speculation for a modular prototyping tool kit for smart jewelry design. ProtoFlakes builds upon the our prior work of Snowflakes  \cite{buruk2021snowflakes}, targeting designers with limited technical expertise with a tool kit to make creative explorations and develop prototypes closely resembling the final products they envision. The design requirements for ProtoFlakes were determined by conducting ideation workshops. From these workshops, we extracted four design parameters that informed the development of the tool kit. ProtoFlakes allows the exploration of form and function in a flexible and modular way and provides a fresh perspective on smart jewelry design. Exploring this emerging area with design speculations informed by ideation workshops has the potential to drive advancements towards more accessible and user-friendly tools for smart jewellery design.

\end{abstract}




\begin{teaserfigure}
\centering
  \includegraphics[width=0.5\textwidth]{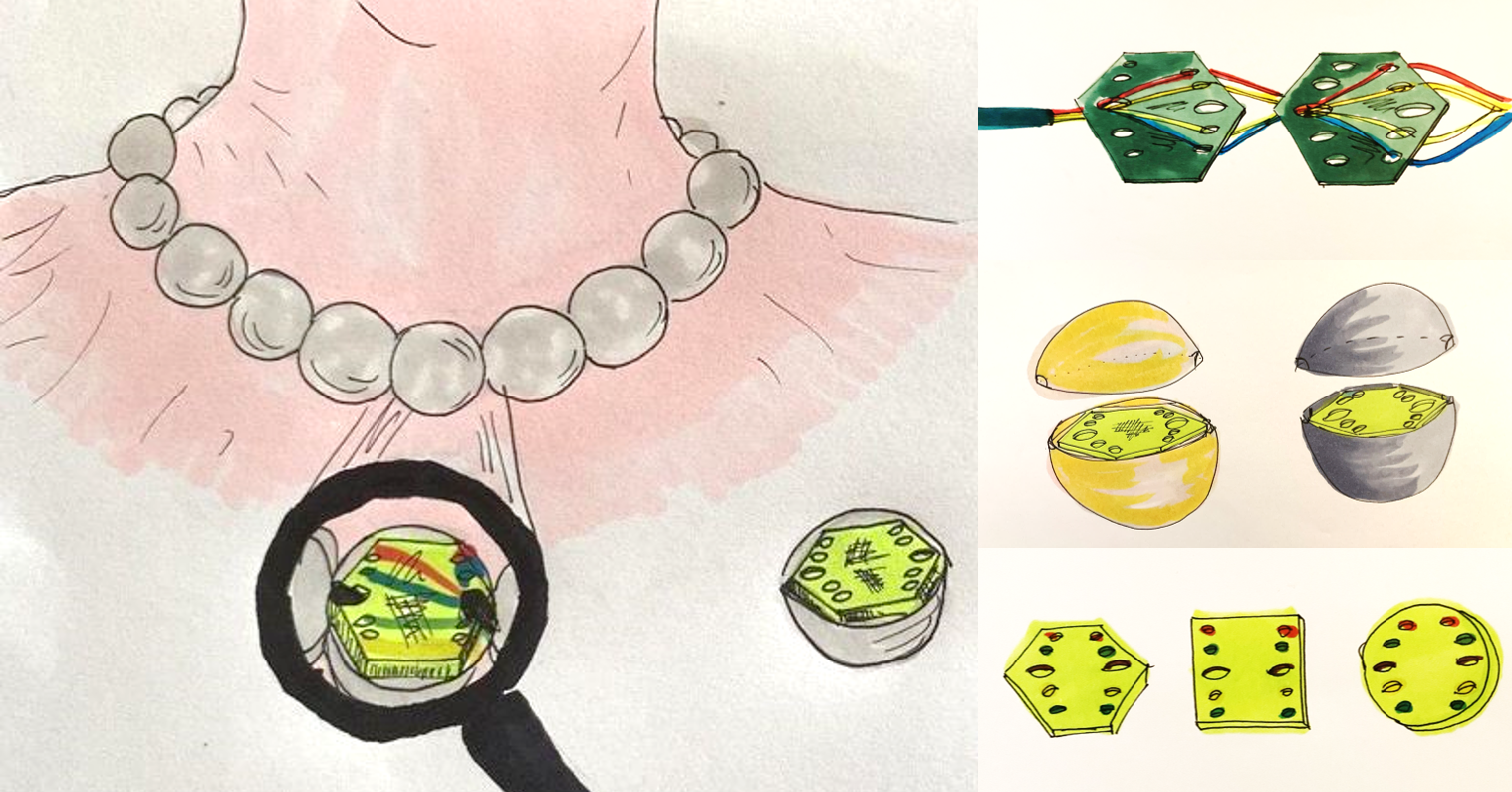}
  \caption{Innovating in smart jewelry design: ProtoFlakes offers a new level of flexibility and customization.}
  \Description{Innovating in smart jewelry design: ProtoFlakes offers a new level of flexibility and customization.}
  \label{fig:teaser}
\end{teaserfigure}


\maketitle

\section{Introduction}

Smart jewelry, a subset of smart wearables, encompasses a diverse range of products such as smart rings, bracelets, necklaces, pendants, glasses, and headwear \cite{song2019innovation}. These products are designed to offer wireless connectivity, sensor and actuator applications, and social interaction features. As a subcategory of jewelry, smart jewelry also falls under the larger umbrella of products where aesthetic expectations are high \cite{insel2018snowflakes}. Therefore, the design of smart jewellery presents a unique set of challenges, as it requires a delicate balance between technical functionality and aesthetic appeal. To address these challenges, it is essential to develop smart jewellery prototyping tool kits that can enable designers to make creative explorations and produce prototypes closely resembling the final products they envision. 

Providing designers with functionality in a modular and flexible fashion through such prototyping kits while allowing the mixing of smart and non-smart components is an ongoing effort in the field of smart jewelry design. One popular approach for achieving modularity is through the use of a skeleton with interconnected modules. The connection between the modules can be electrical (for communication or power) or structural in order to keep the skeleton together. Many tool kits have been developed using this approach, such as the popular LillyPad \cite{buechley2008lilypad}, Flora \cite{AdafruitFlora}, SoftMod \cite{lambrichts2020softmod}, and Snowflakes \cite{buruk2021snowflakes} kits, which all provide modularity to varying extents. 

Another ongoing effort in this emerging field is the seamless integration of functionality with form. Depending on the design decisions and the nature of the target functionality, a designer may choose to highlight or conceal the smart components. BLInG \cite{jo2021bling} is a notable example of blending in and hiding the non-smart components, while tool kits such as SkinKit \cite{ku2021skinkit} and Wearable Bits \cite{jones2020wearable} offer the ability to integrate smart and non-smart components in a visible or hidden manner.

Despite the availability of a wide variety of smart jewelry prototyping tool kits, such as LillyPad \cite{buechley2008lilypad}, Flora \cite{AdafruitFlora}, SoftMod \cite{lambrichts2020softmod} and Snowflakes \cite{buruk2021snowflakes}, there still remains a gap in terms of tools that allow designers to make creative explorations with options to hide, reveal, and/or resize smart components while mixing materials and colors. Snowflakes \cite{buruk2021snowflakes} is a previous work by the authors of this paper, and while it was a significant step forward in this direction, it still has limitations in terms of size and form. To address this gap, we conducted ideation workshops on the form aspect of prototyping smart jewelry with designers and engineers. From these workshops, we extracted four design parameters by analyzing the outputs. Based on these parameters, we propose ProtoFlakes as a design speculation for a tool kit that will allow prototyping smart jewelry that meets the form-related needs of designers while addressing the limitations of Snowflakes.

\section{Design Parameters from Ideation Workshops}

To address the challenges of designing smart jewelry that meets both form and function requirements, particularly for designers with limited technical expertise, we conducted a series of ideation workshops with a diverse group of participants as summarized in Table \ref{tab:participants}. 

\begin{table}[h]
\centering
\begin{tabular}{|c|c|c|}
\hline
\textbf{Group} & \textbf{Designers} & \textbf{Engineers} \\
\hline
1 & 5 & 1 \\
\hline
2 & 3 & 3 \\
\hline
3 & 4 & 0 \\
\hline
\end{tabular}
\caption{Profile of participants who attended the ideation workshops.}
\label{tab:participants}
\end{table}

\begin{figure}[h]
\centering
\includegraphics[width=0.5\textwidth]{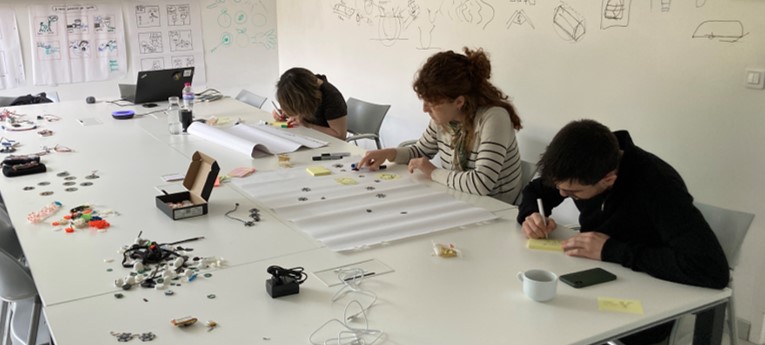}
\caption{Attendees actively engaged in the ideation process during one of our ideation workshops.}
\label{fig:attendees-writing}
\end{figure}

The workshops began with providing background information on prototyping and smart jewelry, followed by hands-on exploration of various prototyping materials, including all versions of Snowflakes  (ranging from the 3D printed speculation version \cite{insel2018snowflakes} to the final fully functional version \cite{buruk2021snowflakes}) and an in-depth explanation of the development process of Snowflakes. After gaining an understanding of the topic, participants were encouraged to generate new ideas and visualize them using the provided materials or through drawings (as in Figure \ref{fig:attendee_drawing}) and written wish statements. As seen in Figure \ref{fig:attendees-writing}, attendees were actively engaged in the ideation process. 

We recorded the entire sessions and, later, transcribed for analysis to extract design parameters. Through this iterative process, we were able to gain valuable insights into the needs and desires of designers working in the field of smart jewelry. During these workshops, we collected 71 wish statements and 24 product ideas from 16 participants. We analyzed the results to extract four design parameters that informed the design of a smart jewelry prototyping tool kit. These parameters include:

\subsection{Variable Connections} 
The ability to easily connect and disconnect various components of the prototype is a crucial aspect of a smart jewelry prototyping tool kit. This ability enables designers to quickly and easily create unique and innovative designs without the need for technical expertise. The ease of connecting and disconnecting different parts of the prototype would enable designers to experiment with different configurations and explore new possibilities. During the workshop, a total of 30 ideas contributed to the formation of this parameter, an example of which can be seen in Figure \ref{fig:attendee_drawings}A. Additionally, Snowflakes tool kit, a modular prototyping tool developed and evaluated in our previous work, served as a reference point for the attendees in terms of connections and ease of use.

\subsection{Grip and Placement}
\begin{figure}[h]
\centering
\includegraphics[width=0.5\textwidth]{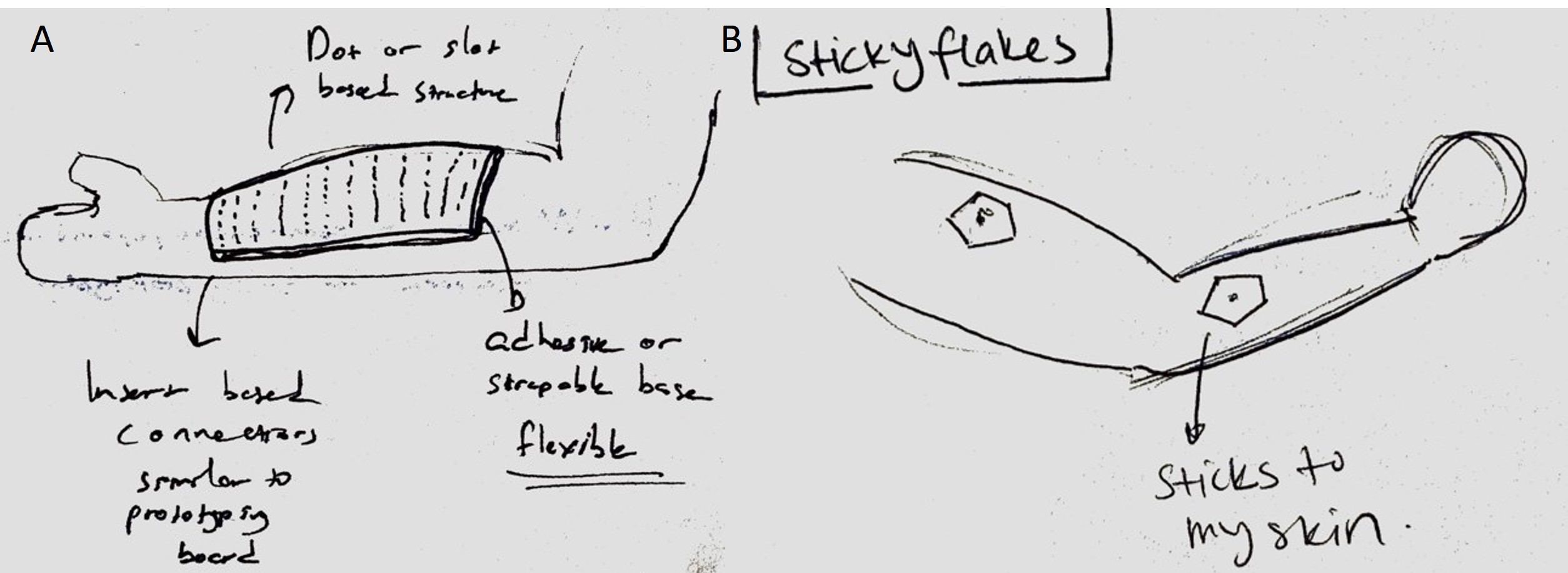}
\caption{A) An example of an attendee's drawing showing their idea about connections (handwritten text: "Dot or slot based structure", "adhesive or strapable base, flexible", "Insert based connections, similar to the prototyping board"). B) An example of an attendee's drawing showing their idea about grip and placement (handwritten text: "Sticklyflakes", "sticks to my skin").}
\label{fig:attendee_drawings}
\end{figure}

Comfortable and secure wearability is a key consideration in smart jewelry design. A prototyping tool kit should provide designers with the ability to easily adjust the placement and grip of the prototype to ensure that it is comfortable and secure for the user to wear. This would enable creation of prototypes that are both functional and comfortable, a crucial factor in user acceptance. This design principle was highlighted during our workshops, where we collected a total of 27 ideas that contributed to the formation of this parameter, an example of which can be seen in Figure \ref{fig:attendee_drawings}B. As previously emphasized in the Snowflakes tool kit \cite{buruk2021snowflakes} , grip and placement is a crucial aspect of smart jewelry prototyping.

\subsection{Size and Shape}
The ability to easily adapt the size and shape of a prototype to align with the form requirements of the design is vital in smart jewelry prototyping. This enables designers to create prototypes that closely resemble the desired final product and facilitates creative explorations. Being able to easily adjust the size and shape of the prototype, designers can create prototypes that better match the user's requirements and aesthetic preferences. During our workshop series, a total of 24 ideas informed the formation of the size and shape parameter.

\subsection{Mixing}
The prototyping tool kit should also enable designers to easily mix different materials and colors in the prototype, allowing for the creation of unique and innovative designs that incorporate both smart and non-smart components, and to adapt the design to the aesthetic requirements of the final product. This feature would provide designers with increased flexibility and creativity in their designs, allowing them to create more sophisticated and personalized prototypes. The parameter for mixing materials and colors was shaped collectively by 19 ideas generated during the workshops.

The proposed tool kit ProtoFlakes aims to enable designers to easily explore different form and function options in smart jewelry design, creating prototypes that closely resemble the desired final product, without the need for technical expertise.
 
\section{Design of ProtoFlakes}
\subsection{Benchmarks and Gap}
In order to design a smart jewelry prototyping tool kit that addresses the identified design parameters, it is important to understand the current state-of-the-art in smart jewelry prototyping tools. We conducted a comprehensive review of existing tool kits such as LillyPad \cite{buechley2008lilypad}, Flora \cite{AdafruitFlora}, SoftMod \cite{lambrichts2020softmod}, littleBits \cite{littlebits}, Wearable Bits \cite{jones2020wearable} and Snowflakes \cite{buruk2021snowflakes}. Our analysis revealed gaps in the ability of current tool kits to address the design parameters of connections, grip and placement, size and shape, and mixing, as can be seen in Table \ref{tab:comparison}. These gaps limit the ability of designers to easily explore different form and function options in their designs, hindering the creative process by limiting experimentation with different materials, connections, grip and placement, and sizes and shapes. 

Our previous work, Snowflakes, provided a size option (two different size options depending on the user preference with the ability to mix different sizes) and hexagons as shape (as it helps creating amorph forms). However, during the development of Snowflakes, a flexible solution to address the issues of providing variable sizes and shapes could not be found. The gap also highlights a need for a tool kit that addresses these specific design parameters, which is what we propose with ProtoFlakes. By addressing this gap, ProtoFlakes aims to provide designers with a more flexible and modular tool kit for prototyping smart jewelry, allowing for greater creative exploration and the potential for more innovative and user-friendly designs. ProtoFlakes addresses this issue by offering a variety of designs that can fit within a minimum size of 8mm diameter.

\begin{table}[h]
\centering
\begin{tabular}{|c|c|c|c|c|}
\hline
  & Var. & Mix. & Var. & Cus. \tabularnewline
  & Con. & Mat. & G.\&P. & S.\&S. \tabularnewline
\hline
Lilypad Flora & yes & no & yes & no \tabularnewline
\hline
littleBits & no & no & no & no \tabularnewline
\hline
Wearable Bits & yes & yes & yes & yes \tabularnewline
\hline
SoftMod & no & no & yes & no \tabularnewline
\hline
Snowflakes & no & yes & yes & no \tabularnewline
\hline
ProtoFlakes & yes & yes & yes & yes (from 8mm of diam.) \tabularnewline
\hline
\end{tabular}
\caption{Comparison of different smart jewelry prototyping tool kits on four design parameters (Var. Con. = Variable Connections, Mix. Mat. = Mixing Materials \& Colors, Var. G. \& P. = Variable Grip \& Placement, Cus. S.\& S. = Customizable Size \& Shape).}
\label{tab:comparison}
\end{table}

\subsection{Proposed Design}
Based on the identified gaps in existing tool kits, we propose the design of ProtoFlakes, a modular and plug-and-play smart jewelry prototyping tool kit. The design of ProtoFlakes is informed by the design parameters of connections, grip and placement, size and shape, and mixing. The modular design allows for easy connection and disconnection of different components, while the plug-and-play functionality enables easy exploration of different form and function options. The tool kit also allows for easy mixing of different materials and colors to adapt the design to the aesthetic requirements of the final product.


ProtoFlakes is a smart jewelry prototyping toolkit that aims to empower designers by providing flexibility in their designs. The key design principle of ProtoFlakes is the compact size of the modules, which are as small as 8mm in diameter and 0.5mm in thickness. This allows designers to incorporate ProtoFlakes modules into small covers, beads or other small spaces, giving them the freedom to create unique and intricate designs without being limited by size constraints. The small size of the modules was determined by considering the smallest possible components required for building electronic boards, such as PCBs and other components, while still ensuring the toolkit is ergonomic and accessible for a wide range of designers.

ProtoFlakes offers four types of modules:
\textbf{1. Sensor and actuator modules}, which include PCB boards with surface mount components such as accelerometers, LEDs, or heart rate sensors;
\textbf{2. Main modules}, which include PCB boards with microcontrollers, and integrated circuits for wireless connection;
\textbf{3. Battery and charging modules}, which include PCB boards with integrated circuits to handle power requirements, surface mount batteries, and interfaces to charge the batteries or provide energy to the system wired or wireless; and \textbf{4. Connection threads}, which provide connection between the modules using the i2c protocol. The connection between the four-pin interface is formed by conductive threads with shielded lengthwise except the endings.

\begin{figure}[h]
\centering
\includegraphics[width=0.5\textwidth]{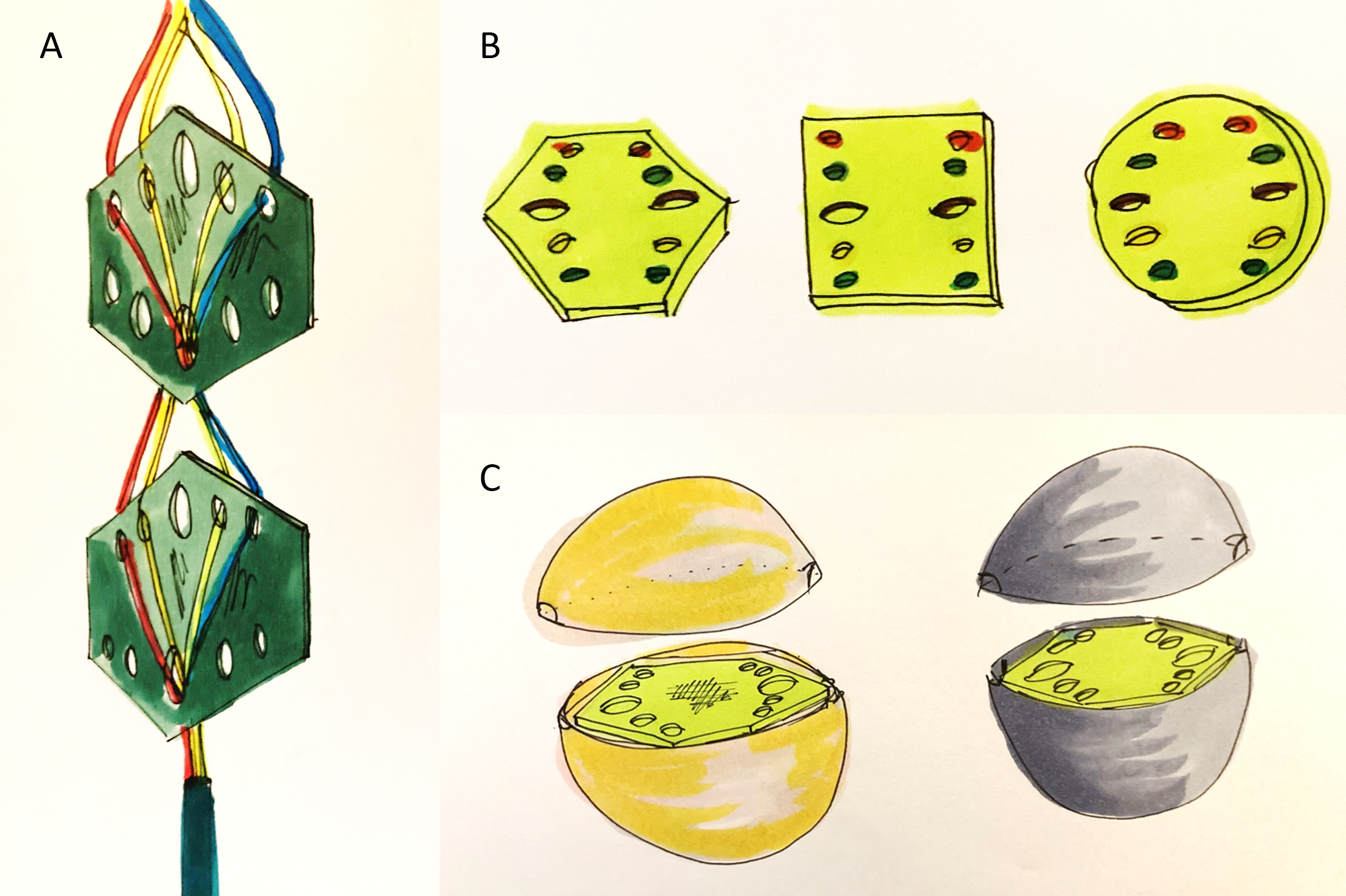}
\caption{A) ProtoFlakes modules in various shapes, B) ProtoFlakes modules connected through vias, C) An example of a prototype using ProtoFlakes with different cover materials and colors, such as golden and silver beads.}
\label{fig:features}
\end{figure}

ProtoFlakes offers various shape options to designers, one of which is hexagons as a tribute to our former work Snowflakes. The PCBs can have various shapes such as hexagons, triangles, squares and more, (see Figure \ref{fig:features}A), and have vias (holes with conductive coatings) providing connection points for threads. As long as all the vias of the same type are connected between the modules, the circuit will be completed and working (see Figure \ref{fig:features}B). Furthermore, ProtoFlakes offers greater flexibility in terms of shape and size compared to our previous work Snowflakes.

As the modules are as small as 8mm in diameter, they can be covered with cases of comparable sizes, allowing designers to create unique designs such as a pearl necklace with hidden sensors and actuators within the beads. The covers can be made of any material or color, allowing designers to mix and match different materials, such as golden and silver beads as shown in Figure \ref{fig:features}C. 

\begin{figure}[h]
\centering
\includegraphics[width=0.5\textwidth]{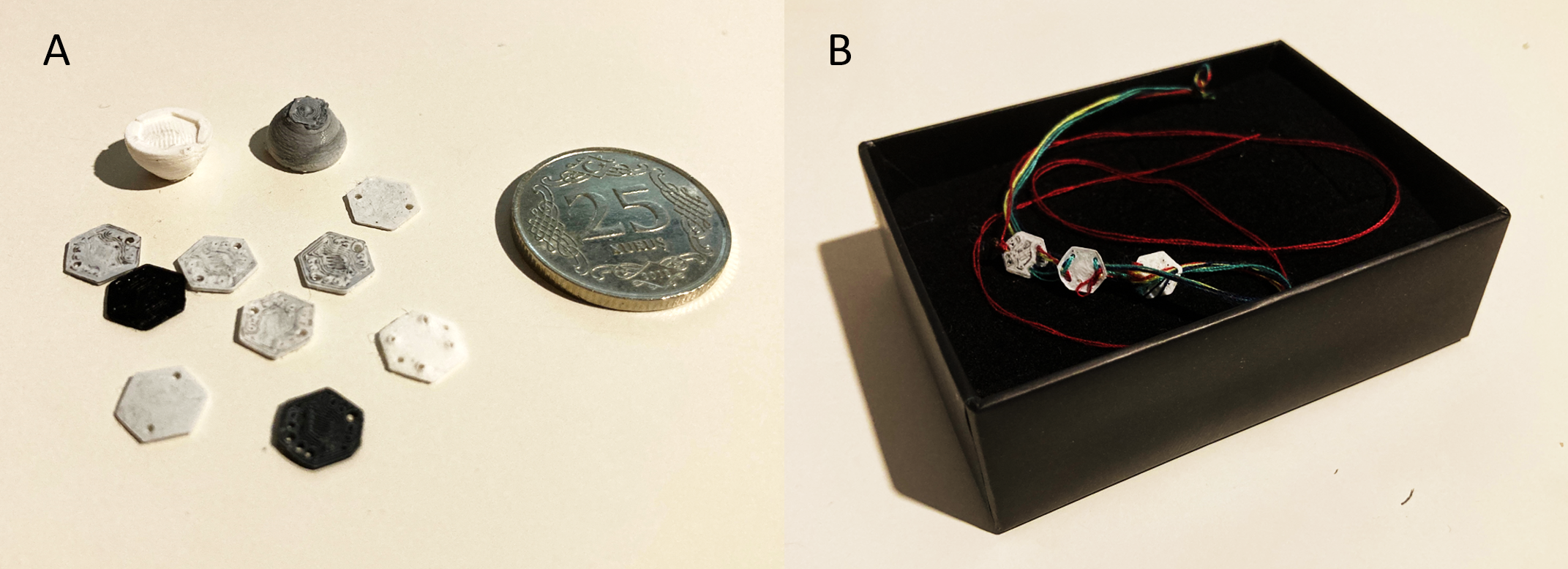}
\caption{A) An example of 3D printed ProtoFlakes modules with a coin to depict the size, B) An example of a bracelet made with 3D printed ProtoFlakes modules showing connections with conductive threads.}
\label{fig:3d_printed}
\end{figure}

To illustrate the ergonomics and overall concept of the tool kit, we have created 3D printed conceptual modules of ProtoFlakes. Figure \ref{fig:3d_printed}A shows a 3D printed ProtoFlakes module next to a coin, to depict the size of the module. Figure \ref{fig:3d_printed}B shows a bracelet made with 3D printed ProtoFlakes modules

Based on the design parameters identified, we created ProtoFlakes as a tool kit for smart jewelry prototyping. The ability of ProtoFlakes to meet the form-related needs of designers will be assessed through additional testing and feedback from target users.

\section{Implementation Plan}

\subsection{Hardware}
ProtoFlakes will consist of various hardware components such as microcontroller boards, sensors, actuators, and connectors. These components will be selected based on their compatibility with the proposed design of ProtoFlakes and their ability to address the design parameters. The hardware components will be sourced from reliable suppliers and will undergo testing to ensure their performance and durability.

To ensure that all components fit within the 8mm diameter size constraint, a list of potential options have been researched and compiled in Table \ref{table:hardware_options}. This includes options for a microcontroller, wireless communication, accelerometer, heart rate sensor, LED, SMD supercapacitor, and charging IC.

\begin{table}[h]
\centering
\renewcommand{\arraystretch}{1}
\begin{tabular}{|c|c|}
\hline
\centering \textbf{Component} & \centering \textbf{Model} \tabularnewline
\hline
\centering Microcontroller & \centering Atmel ATSAMD21G18A-MU \tabularnewline
\hline
\centering Wireless & \centering Texas Instruments CC2640R2F \tabularnewline
\hline
\centering Accelerometer & \centering Bosch Sensortec BMI160 \tabularnewline
\hline
\centering Heart Rate Sensor & \centering Maxim MAX30101  \tabularnewline
\hline
\centering LED & \centering Kingbright WP7113SGC \tabularnewline
\hline
\centering SMD Battery & \centering Murata GRM188R71H104KA93D \tabularnewline
\hline
\end{tabular}
\caption{A selection hardware options for ProtoFlakes}
\label{table:hardware_options}
\end{table}


It's worth noting that the models provided are examples and there may be other options available that also meet the size requirements of 8mm diameter. Additionally, the notes provided are brief and more detailed information about each component can be found on the respective datasheets.

\subsection{Software}
The software development for ProtoFlakes will focus on creating a user-friendly interface and ensuring the compatibility of the software with the hardware components. The software will include features such as a drag-and-drop interface for easy connection and disconnection of different parts, a library of pre-designed forms and functions, and the ability to customize and save designs. The software will be developed using open-source platforms and will be continuously updated and improved based on user feedback.

\section{Limitations}

ProtoFlakes is currently in the iterative implementation phase and has not yet been fully implemented or tested.

\subsection{Technical Limitations}
ProtoFlakes, like any other tool kit, has some technical limitations. For example:

\begin{itemize}
\item The tool kit may have limited compatibility with certain types of sensors and actuators.
\item The tool kit may also have limitations in terms of the complexity of designs it can handle, and the number of simultaneous connections it can support.
\item The limitations of the technology used in ProtoFlakes, such as battery life and communication range, may affect the overall performance and usability of the tool kit.
\item The scalability of ProtoFlakes and its ability to be used for larger projects or with more complex designs is also a limitation that needs to be considered.
\end{itemize}

\subsection{Design Limitations}
ProtoFlakes is designed to address certain design parameters, however, it may not be suitable for certain types of designs. For example, the tool kit may not be suitable for designs that require a high level of precision or designs that require a large number of connections. ProtoFlakes design is based on the results of the ideation workshops and the benchmarking study, but it may not meet the needs of all designers or all types of jewelry designs. The sample size for the workshops is limited and may not be representative of the entire target audience and the scope of the research is limited to the design of a tool kit for smart jewelry prototyping, and does not include the commercialization or market analysis of the product.

\section{Conclusion}
In this paper, we presented ProtoFlakes, a design speculation for a modular prototyping tool kit to design smart jewelry. Building on our previous work Snowflakes, we conducted ideation workshops with designers and engineers to extract design parameters and proposed ProtoFlakes as a tool to address these parameters. Through this research, we aim to contribute to the field of smart jewelry design by providing a tool that addresses the needs of designers with limited technical expertise.

ProtoFlakes is not a tool to design a final product, but rather a tool to provide a roadmap for designers before finalizing their designs. It allows designers to explore form and function options in a flexible and modular way. The tool kit provides a background and a base for designers to have a fresh and new look at smart jewelry design.

It's important to mention that ProtoFlakes is a speculation and further studies are needed to validate its potential. However, this new area in the field is worth studying as it could contribute to the development of more accessible and user-friendly tools for smart jewelry design. In future studies, we plan to further evaluate ProtoFlakes through user testing and make any necessary adjustments to the design. We also plan to explore the potential of ProtoFlakes in other areas of smart wearables design.

With ProtoFlakes, we believe that designers will be able to develop an enormous number of jewelry designs that will help reveal the potential of smart jewelry in our daily lives. Despite the rapid growth of smart wearables, smart jewelry has yet to fully penetrate our lives as much as other smart devices. With ProtoFlakes, we believe that designers will be able to find new applications of smart jewelry to understand and create user needs and solve them. By providing access to a wide variety of designers without the need for programming or hardware knowledge, we believe that a breakthrough may happen in the industry around smart jewelry and they may become an essential part of our daily lives.

Having ProtoFlakes penetrate into the design community may also eventually expand the community to everyone and create an essential step on the journey of mass customization. As the smaller modules become possible and accessible, creating any device or experience becomes also possible for any person.

In conclusion, the field of smart jewelry is an emerging and rapidly growing area of smart wearables that includes a wide range of products such as smart rings, bracelets, necklaces and more. Despite the development of tool kits that aim to provide designers with functionality in a modular and flexible fashion, there remains a gap in the area of smart jewelry prototyping tool kits that allow designers to make creative explorations while mixing materials and hiding, revealing, and/or resizing smart parts. By addressing this gap, we can expect to see more innovation and unique designs in the field of smart jewelry. Our proposed ProtoFlakes tool kit aims to bridge this gap and provide designers with more options to create truly unique and innovative designs. With the help of ProtoFlakes, we can anticipate a surge in the number of successful and form-related smart jewelry prototypes, and thus, an increase in the quality and diversity of smart jewelry products that meet the aesthetic expectations of users.

\bibliographystyle{ACM-Reference-Format}
\bibliography{sample-base}










\end{document}